# Simulations of the consequences of tongue surgery on tongue mobility: Implications for speech production in post-surgery conditions.

Stéphanie Buchaillard[1], Muriel Brix[2], Pascal Perrier[1] & Yohan Payan[3]

[1]ICP/GIPSA-lab, UMR CNRS 5216, INP Grenoble, France

[2]Service de Chirurgie Maxillo-faciale, Grenoble University Hospital, Grenoble, France

[3]TIMC-IMAG, UMR CNRS 5525, Université Joseph Fourier, Grenoble, France

Corresponding author:

Pascal Perrier
ICP/GIPSA-lab
INPG, 46 Avenue Félix Viallet
38031 Grenoble cédex 01
Pascal.Perrier@gipsa-lab.inpg.fr

# ABSTRACT

**Background** In this paper, we study the ability of a 3D biomechanical model of the oral cavity to predict the consequences of tongue surgery on tongue movements, according to the size and location of the tissue loss and the nature of the flap used by the surgeon.

**Method** The core of our model consists of a 3D biomechanical model representing the tongue as a Finite Element Structure with hexahedral elements and hyperelastic properties, in which muscles are represented by specific subsets of elements. This model is inserted in the oral cavity including jaw, palate and pharyngeal walls. Hemiglossectomy and large resection of the mouth floor are simulated by removing the elements corresponding to the tissue losses. Three kinds of reconstruction are modelled, assuming flaps with low, medium or high stiffness..

**Results** The consequences of these different surgical treatments during the activations of some of the main tongue muscles are shown. Differences in global 3D tongue shape and in velocity patterns are evaluated and interpreted in terms of their potential impact on speech articulation. These simulations have shown to be efficient in accounting for some of the clinically observed consequences of tongue surgery.

**Conclusion** Further improvements still need to be done before being able to generate easily patient-specific models and to decrease significantly the computation time. However, this approach should represent a significant improvement in planning tongue surgery systems and should be a very useful means of improving the understanding of muscle behaviour after partial resection.

# I. Introduction

Tongue surgery can have severe consequences on tongue mobility and tongue deformation capabilities. It can generate strong impairments of three basic functions of the human life, namely masticating, swallowing and speaking, which induce a noticeable decrease of the quality of life of the patients. Several assessments of these impairments have been done in post-surgery conditions on patients, who had undergone tongue and/or mouth floor carcinoma resections and reconstructions. These assessments were based either on well-established functional protocols, such as the *University of Washington Quality of Life Instrument* (1), evaluating among other tasks the patients' communication skills (2-3), on specific acoustic speech corpora (4-5) assessing the lost of speech movements amplitude and accuracy, or on quantitative measurements of the oral cavity (6-7). Statistical analyses of the results collected on large enough samples of patients could provide valuable indications about the impacts of surgery and reconstruction techniques

In the line of works carried out in predictive medicine to set up systems of computer aided surgery, such as in orthopaedic (spine and pelvis) (8-10), cranio-facial (11-13) surgeries, this paper presents the current achievements of a long term project aiming at predicting and assessing the impact of tongue and mouth floor surgery on speech production. The ultimate objective of this project is the design of a software with which surgeons should be able 1) to design a 3D biomechanical model of the tongue and of the mouth floor that matches the anatomical characteristics of each patient specific oral cavity, 2) to simulate the anatomical changes induced by the surgery and the possible reconstruction, and 3) to quantitatively predict and assess the consequences of these anatomical changes on tongue mobility and speech production after surgery.

The interest of this approach is linked with functional aspects of the surgical reconstruction. Depending on the site and the size of the tissue losses, the functional alteration can originate from a lack of surface, lining and texture, a lack of volume and bulk, or a lack of mobility. The choice of the flap used for the reconstruction is also an important issue. From this perspective, a biomechanical model of the tongue is helpful, since it allows simulating the planned surgery in terms of cut design. Such a model could also be helpful to select a specific kind of flap in comparison to others, since it allows a quantitative study of the mobility of the tongue for different reconstructions.

The 3D model of the oral cavity is presented in the "Material and Methods" section, together with a description of typical examples of simulations of tongue resection and reconstruction. In the "Results" section the consequences of the simulated surgeries on tongue displacement are shown, and an analysis of their potential consequences for speech production is proposed. Perspectives and further developments of the projects are discussed in the "Conclusion" section.

## II. Material and method

Our work is based on four basic steps:

. Development of a 3D biomechanical model of the oral cavity

. Implementation on this model of the anatomical and mechanical transformations generated by the most frequent tongue resections and their associated reconstructions.

. Simulations with the model of the displacements and the shaping of the tongue associated with the activation of the main tongue muscles, before and after surgery.

. Interpretation of these simulations in terms of implications for speech production after tongue surgery.

The 3D oral cavity model was originally designed by Gérard *et al.* (14-15) and it was further enhanced recently for speech production control by Buchaillard *et al.* (16). It was developed in the ANSYS$^{TM}$ Finite Elements software environment. Models of the jaw, the teeth, the palate, the pharynx, the tongue and the hyoid bone are its basic constituents. The tongue and the hyoid bone are represented by volumes meshed with 3D hexahedral and tetrahedral elements respectively, while the other structures are modelled by 3D surface elements. These surface elements are essentially considered as oral cavity limits with which the tongue interacts due to mechanical contacts.

### *II.1.    The 3D biomechanical model of the tongue*

#### II.1.1. Some basics about tongue anatomy

Among the nine muscles that act on the tongue structure, there are five extrinsic muscles that originate from structures external to the tongue (mostly bony structures) and insert into the tongue: the genioglossus, the styloglossus, the hyoglossus, the geniohyoid and the palatoglossus. Four additional intrinsic muscles are completely embedded in the tongue structure: the inferior and superior longitudinal, the verticalis and the transversus. For all these muscles the fibres are mainly oriented in the front/back or in the vertical direction, except for the transversus, which fibres are transversally oriented. In addition the majority of them are paired muscles, symmetrically distributed on each side of the midsagittal plane of the head. Detailed descriptions of tongue anatomy can be found in (17), (18), and (19).

As concerns the functions of these muscles for tongue moving and tongue shaping, the description in this paper will be limited to the muscles that have been used in the simulations shown in section III. The anterior bundle of the genioglossus tracts the apex down and forward. Its middle bundle propels the tongue out of the oral cavity. Its inferior and posterior bundles tracts the base forward and elevates the core of the tongue. The styloglossus raises the base, and elevates globally the tongue in the velar area. The hyoglossus lowers the tongue and brings it back after propulsion. The superior longitudinal shortens the tongue and tracts the apex up and back. The transversus shortens the transversal diameter and makes the tongue sharper.

The floor of the mouth is a crescent-shaped space extending from the lower alveolar ridge to the ventral surface of the tongue. Its anterior part allows good mobility and protraction of the tongue.

### II.1.2. Finite elements mesh design

The mesh design process is completely described in (14) and (20). The main guidelines are the following. In a first step, a generic mesh was designed from the Visible Human Project TM data set. It specifies the shape of the elements taking into account the main trends of muscle anatomy such as the main orientations of muscles fibres. A precise implementation of muscles was done in this generic mesh thanks to accurate anatomical data extracted from (17), (18), and (19) (see (14) for a comprehensive description). This approach allowed a realistic description of the muscular anatomy of the tongue. However, since it is based on data collected from a frozen cadaver, the geometry of the tongue is likely to be altered as compared to living subjects.

This is why, in a second step, the geometry of the generic mesh was transformed, so that it matches the external 3D contours of the tongue of a living subject (PB), for whom many different kinds of data (CT-Scans, MRI data, X-Ray data) about the head and the oral cavity had been collected. The external tongue contours were extracted from the MRI data, with manual image segmentation (21).

Thus, a new subject specific finite element mesh was achieved via a mesh-matching algorithm (22) projecting the external nodes of the generic mesh onto the tongue surface measured from PB's MRI data, followed by an interpolation process specifying the internal nodes positions from the new external mesh while respecting the original structure of the generic mesh. Finally a geometrical refinement was carried out from an X-Ray midsagittal view of PB's oral cavity at rest in a seating position (Figure 1, left panel). This was done because the supine position of the subject during the MRI data acquisition is likely to influence the tongue shape due to gravity. This final adaptation guarantees the best possible definition of the tongue geometry at rest. This mesh was then included in a whole oral cavity by inserting it in the jaw, palate, and pharyngeal surface meshes that were extracted from PB's MRI and CT-scans data (Figure 1, right panel).

### II.1.3. Muscle representation

Muscles are represented as a set of adjacent elements in the tongue mesh. Figure 2 shows the arrangement of muscles activated in section III. It should be noted that the genioglossus has been divided in its three functional parts: posterior, medium and anterior. Macro-fibres, that represent bundles of muscular fibres, were defined by joining finite elements nodes for every muscle, according to the fibres known orientation (represented in red on Figure 2).

Muscle activation is produced by an external generator based on a functional model of muscle force generation mechanisms, the *Equilibrium Point Hypothesis* also called $\lambda$ model (23). The implementation of this model was adapted to the tongue in order to ensure positioning accuracy and stability. Details can be found in (16).

### II.1.4. Tongue biomechanics

The tongue was modelled as a soft tissue attached to the hyoid bone, the jaw and the styloid process, in interaction through soft contacts with the anterior part of the palate and the lower teeth. It consists of a 3D Finite Element model assuming a large deformation framework with a non-linear hyper-elastic constitutive law (15; 24). Arguing the fact that the constitutive law (i.e. the stress-strain relationship) proposed by Gérard *et al*. (24) was extracted from data collected on tongue tissues removed from a fresh cadaver, we proposed to adapt this original law to account for differences between passive tissues and tissues belonging to active muscles. Therefore, two different constitutive equations were introduced: one describing the passive behaviour of the tongue tissues and another one modelling the stiffness of the tissues as an increasing function of tongue muscles activations. For a given element of the model, the passive (respectively the active) constitutive law is used if the element belongs to a region that is supposed to be passive (respectively active). The passive constitutive law was directly derived from this non-linear law (24). However, as the stiffness of passive tissues removed from a cadaver is known to be lower than the one measured on in vivo tissues, it was decided to modify the constitutive law originally proposed in our tongue model. To our knowledge, one of the most relevant in vivo measurements provided in the literature for human muscle stiffness is the one proposed by Duck (25). This work provided a value of 6.2kPa for a human muscle in its rest position, and a value of 110kPa for the same muscle when it was contracted. As the original constitutive law of Gérard *et al*. (15) provided a value of 1.15kPa for the stiffness at the origin (i.e. the Young modulus value at low strains), it was decided to multiply the original constitutive law by a factor of 5.4, in order to reach the 6.2kPa value at the origin while maintaining the overall non-linear shape of the passive constitutive law. Figure 3 plots the corresponding law.

As concerns the constitutive law that describes the mechanical behaviour of an element belonging to an active muscular region, it was decided to arbitrarily multiply the passive constitutive law by a factor that is a function of the muscle activation. The underlying idea is that an activation of the muscle leads to an increase of its stiffness. The multiplying factor was chosen in order to maintain the stiffness value below 110kPa when maximal muscle activation is simulated (25).

The tongue mass density was chosen equal to 1040kg/m$^3$. With the proposed finite element mesh the tongue mass was calculated equal to 105g. The damping was defined by specifying mass and stiffness Rayleigh damping constants that were set respectively to 10 and 0.7.

### II.1.5. Hyoid bone biomechanics

The hyoid bone is represented by 4-nodes 3D tetrahedral elements. Its mass approximates 5g. Its body gives insertion to the mylohyoid, the geniohyoid and the posterior part of the genioglossus, and the greater cornua to the hyoglossus. Nodes on the tongue mesh corresponding to muscle insertions were selected as insertion nodes. A set of 10 springs emerging from the hyoid bone were used to represent the anterior part of the digastric, its posterior part, as well as the sternohyoid, omohyoid and thyrohyoid muscles. The spring's stiffness was set to 200N.m$^{-1}$. The insertion nodes and the hyoid bone define a rigid region to ensure a realistic coupling between the hyoid bone and the surrounding muscles.

### II.1.6. Contacts between tongue and hard structures

Three contact regions were determined to represent the contact between the tongue and 1) the hard palate, 2) the soft palate, 3) the lower teeth and the inner face of the mandible. For each region, a group of elements belonging to the tongue surface, called the contact surface, and to the second region of interest, called the target surface, were selected. At each step of the resolution, collisions and contacts were detected for every surface/target pair and resolved using an augmented Lagrangian contact algorithm (iterative penalty method). We used a contact stiffness factor of 0.10 (ANSYS$^{TM}$ FKN parameter) for the different pairs, and an allowable penetration (ANSYS$^{TM}$ FTOLN parameters) of 0.2mm for the first two pairs (contact with the palate), and 0.3mm for the last one.

## *II.2.* *Simulations of tongue surgery*

Clinically, tongue surgery is essentially considered in two cases: functional disabilities associated with an excess of the tongue volume (macroglossectomy) and presence of carcinomas within the tongue tissues. In some cases the tissues loss due to the resection is compensated by a reconstruction procedure using flaps. In this section simulations of three very common tongue resections used to remove tongue tumours are presented. Some cases of reconstructions have also been simulated.

### II.2.1. Hemiglossectomy

Most of the cancers occur on the lateral border of the tongue at the junction between the middle and posterior thirds. Surgical resection for oral tongue cancer can be aggressive with planned margins of at least 1 cm. For tumours with diameters as large as 2 cm the hemiglossectomy is the recommended surgical treatment. The first surgery that was simulated on the model corresponds to this case. It is shown on Figure 4 for a left hemiglossectomy.

In this implementation, the resection affects the large majority of the tongue muscles. The left part of the styloglossus is removed as well as the left anterior parts of the inferior and superior longitudinal muscles, the left anterior parts of the transverse and vertical muscles and the upper part of the left hyoglossus. The medium and anterior parts of the left genioglossus are nearly entirely removed, whereas its posterior part is only partially affected. Since the resection affects only half the tongue's body, we can expect highly dissymmetric movement of the tongue after such a resection.

### II.2.2. Mouth floor resection

In the anterior floor of the mouth, the smallest tumours involve mucosa. The resection extends then in the mucosa behind the alveolar ridge, to the excluded ventral face of the tongue. This corresponds to the second example of surgery that was implemented on the model (see Figure 5). In that case, the mobile tongue is totally preserved. The anterior part of the genioglossus is removed as well as the totality of two major muscles of the mouth floor, namely the geniohyoid and the mylohyoid muscles. For the moment, we consider surgery that affects the muscles only, not the lingual nerve or the mandible.

### II.2.3. Reconstruction

Reconstruction of the tongue can be achieved with primary closure and local, regional or distant flaps. Resections of the anterior part of mouth floor due to small tumours can be repaired with local flaps. The most common local flap is the buccinator muscle flap. It is versatile and its elastic characteristics seem to be comparable with those of the oral mucosa. The size of the flap is small. Two flaps (one of each side) can be used to repair a defect of the mouth floor. In such cases, the reconstruction allows a better mobility and avoids a spontaneous protrusion of the tongue out of the oral cavity. When the resection includes the ventral tongue or/and the posterior part of the tongue, the reconstruction has to be composite, with a musculocutaneous flap (pectoralis major and latissimus dorsi) or, even better, with an osteocutaneous flap (fibula). These flaps are less versatile and their elasticity significantly differs from the one of the oral mucosa. After surgery, radiotherapy is often necessary, especially when the lymphatic nodes or the bone are involved by the tumour. The surface characteristics of the soft tissues are modified by the radiations and they loose their softness. The final result differs according to whether the patient underwent or not radiotherapy.

For the two modelled resections (namely hemiglossectomy and mouth floor resection), four cases were studied that attempt at reproducing these different characteristics of the flap, and the consequences of a possible radiotherapy. First we simply implemented a resection of the tongue without reconstruction. Then three different kinds of flaps were used : a flap having exactly the same biomechanical properties as the passive tissues of a healthy tongue (see Figure 3), a flap with a stiffness 5 times smaller and a third one with a stiffness 6 times higher. These flaps are totally inactive during the tongue activation.

## III. Results

All the simulations presented below correspond to muscle activation patterns lasting 150 ms.

### *III.1. Asymmetrical movements of the tongue following a hemiglossectomy*

In normal conditions, the tongue movements are coarsely symmetric (pair muscles are activated simultaneously according to a similar pattern of activation). After a hemiglossectomy, this symmetry is broken, since half the tongue is removed and possibly reconstructed, and it induces asymmetries in tongue displacements and shaping. The induced deviations from normal symmetrical movements are more or less important depending on the muscles that are activated. Figure 6 shows a case of important deviation with a frontal view of the tongue mesh after a 150 ms activation of the styloglossus. As expected, in the absence of surgery, this activation pulls the tongue upwards and backwards, and lets the model perfectly symmetric (normal model, Figure 6(a)). When the left-side of the tongue is removed and reconstructed (Figure 6(b-d), an asymmetry of the tongue shaping is observed, characterized by a torsion of the tongue apex and of the tongue dorsum, and by an important deviation of the apex from the midsagittal plane. The physical properties of the flap have a strong impact on the amplitude of the tongue torsion and consequently on the maximal elevation of the tongue in the velar region. Figure 7 (left panel) shows a frontal view of the apical tongue for the same simulations. The torsion decreases when the stiffness of the flap increases. Hence, it seems more difficult to control the positioning of the surface of the tongue when the stiffness of the flap is small. However, these different images show that the stiffness of the flap has nearly no influence on the lateral deviation of the apex (in the three conditions, the central node deviates of approximately 5 mm towards the right). However, Figure 6(b-d) shows that increasing the stiffness of the flap reduces the maximal elevation of the right part of the tongue dorsum, which is consistent with the reduction of the torsion. Figure 7 (right panel) shows a complementary view in the midsagittal plane. Due to the fact that that only the right styloglossus can be activated in reconstructed tongues, the backward and upward displacement of the tongue is reduced compared to the normal condition. It was also observed that the velocity of tongue movements is much lower. We can also notice that, in the midsagittal plane, the tongue shapes associated with the three different flap stiffness differ essentially in the apical area, not in the posterior region.

## III.2. Impacts movements of the tongue after a mouth floor resection

The resection of the mouth floor only affects the posterior genioglossus, the mylohyoid and the geniohyoid. Therefore, consequences can be expected on the ability of the patients to move simultaneously their tongue forwards and upwards. They are simulated below in section III.2.2 for different hypotheses about muscle force recovery after resection. Interestingly, difficulties have also been observed for upward movements in the velar region, which essentially involve the activation of the styloglossus that is not directly affected by the surgery. In order to understand the origins of this observed impairment, the action of this muscle assuming a resection of a medium size tumour of the mouth floor was simulated. Results are described in section III.2.1.

### III.2.1. Effects on tongue elevation in the velar region.

The consequences of a styloglossus activation were evaluated assuming a resection of a medium size tumour of the mouth floor and for reconstructions involving three levels of stiffness of the flaps. Figure 8 shows the resulting shapes of the tongue in the midsagittal plane after activation of the styloglossus. Differences are very small. However, looking at the kinematic characteristics of the movement from the rest position to the final tongue shape, it was observed that velocity is affected by flap stiffness. Table 1 shows the maximal velocity reached during the movement by nodes located on the surface of the tongue: a high stiffness flap reduces the maximal velocity of the tongue (about 20% of decrease) as compared to the intact tongue. Thus, it can be assumed that difficulties in the production of upward movements in the velar region could be the result of a too high stiffness of the flap. Moreover, as illustrated in table 1, using softer flap could limit the disturbances caused by the resection and the reconstruction for this particular movement.

### III.2.2. Effect on tongue advancement

The last series of tests presented in Figure 9 shows the co-contraction of the posterior genioglossus, styloglossus and transversus, resulting in a forward and upward movement of the tongue for a normal configuration. During the resection of the floor of the mouth, only the anterior part of the posterior genioglossus is removed, resulting in a shorter muscle which fibres insert on the flap after the reconstruction. Little is known about the way muscles contract once they have been partially shortened, and the level of activation that can be expected. Three options were tested for the activation of the fibres of the posterior genioglossus: 1) absence of activation, 2) low level of activation compared to the normal case, 3) similar level of activation compared to the normal case.

Results of the simulations show how difficult it can be to pull the tongue upward and front, once the genioglossus has been partially removed, even in case 3 where it is hypothesized that muscle activation is preserved. Once the genioglossus insertions on the superior mental spine of the mandible have been removed, it losses its capacity to protrude and elevate the tongue. Increasing the stiffness of the flap facilitates clearly the tongue propulsion, even if it is still not sufficient to retrieve the same shape as in the normal case.

These results show that in case of mouth floor resection and reconstruction the stiffness of the flap can dramatically influence tongue mobility. However, while it seems to be better to use soft flap to preserve tongue elevation in the velar region, stiff flaps appear to be better to maintain the front/up movement of the tongue. This shows how interesting the use of a biomechanical model of the tongue can be in the planning of tongue surgery in order to limit the negative effects of the surgery in terms of tongue mobility.

### *III.3.     Interpretation in terms of impact on speech production*

The control of the tongue for speech production essentially consists of an accurate positioning and shaping of this articulator in reference to the external walls of the oral cavity, namely the hard palate, the soft palate and the pharyngeal wall. This positioning determines first the size of the different resonance cavities of the vocal tract and their coupling, and it influences the spectral characteristics of the sound. It also determines the aeroacoustic nature of the airflow, laminar (when tongue is not too close to the external walls) or turbulent (in case of small cross-sectional area of the oral cavity), which differentiates vowels from consonants. Time characteristics are also very important in speech production, which means that the velocity of the articulatory movements has to match a number of constraints.

It was shown above that a hemiglossectomy generates a strong deviation of the tongue positioning in the coronal plane. Consequently, in the absence of reconstruction, new strategies have to be developed by the patients in order to compensate for this deviation. Further works with the model should tell us whether such compensations are actually possible. More specifically, an important issue is whether the residual volume of the tongue in the palatal region is large enough to allow the production of the oral cavity closure that is required for the production of turbulent airflows underlying the production of alveolar and palatal consonants. In case of reconstruction, the volume should not be a problem any more. In this case it was shown that the control of the surface of the tongue, and then of the cross-sectional shape of the oral cavity (important for the characteristics of turbulent airflows) was easier with stiff flaps. However, it was also shown that reconstruction involving stiff flaps strongly decrease the amplitude of upward tongue movements. As a consequence, the positioning of the tongue near the hard and the soft palates is largely impaired, inducing difficulties in the production of high vowels such as /i/, /e/, /y/ or /u/, or of consonants such as /t/ or /k/. This difficulty was clearly observed in former studies of patients who underwent hemiglossectomy (5).

In the same study (5), it was observed that patients having undergone an important mouth floor resection often had strong difficulties to produce sounds articulated in the velar region such as vowel /u/ or consonant /k/. The simulations shown above in section III.2.2 provide an interesting explanation for this phenomenon: the stiffness of the flap was probably too high. Indeed, in these conditions, according to our simulations, even if these patients could have been able to produce the right positioning and shaping of the tongue, the movement velocity could have decreased so much that it would not longer be possible to shape properly the tongue within a speech sequence with the correct timing. At the same time, our simulations have also shown that in case of mouth floor reconstruction, stiff flaps would allow more forward displacement, facilitating the production of high front vowels such as /i/, /y/ and /e/.

Hence, the simulations presented in this paper are in quite good agreement with observations made on patients after tongue surgery. They also show how useful such simulations can be for surgery planning, since they demonstrate that in case of reconstruction of the moving part of the tongue, the stiffness of the flap has to be very carefully taken into consideration.

# IV.   Conclusion

Simulations with a realistic 3D biomechanical model of the tongue have shown that some of the clinically observed consequences of tongue surgery, including partial resection and reconstruction with a flap, on tongue mobility and speech articulation could be well accounted for by such a model. The potential role of the mechanical characteristics of the flap on tongue mobility was shown. Moreover, it appeared that controlling flap stiffness could help preserving speech production capabilities.

As shown in section V.2.2. our approach can also be very useful to improve the understanding of muscle behaviour after partial resection. Comparison of simulation results with data collected on patients could shed light on the hypothesis (no activation, partial activation or full activation) that seems to be the most realistic.

Therefore, using such a model should represent a significant improvement in planning tongue surgery systems, in order to preserve as much as possible tongue mobility during speech production, and therefore the patients' Quality of Life.

Further improvements of the model have to be done before these ultimate objectives can be actually reached. First of all, fast finite element algorithms have to be implemented and tested in order to significantly decrease the computation time, and to reach simulation durations compatible with an interactive use of the model. First improvements in this direction have been recently done (26-27). Mesh-matching algorithms in the line of those developed by Couteau *et al* (22) have also to be elaborated in order to allow the surgeons to adapt the generic tongue model to the anatomy of each specific patient.

## Acknowledgments:

This work is supported by a grant of the Region Rhône-Alpes (Emergence Program).

The authors express their grateful thanks to Pierre Badin (ICP/GIPSA-lab) for sharing its MRI and X-Ray data, to Matthieu Chabanas (ICP/GIPSA-lab) for advices and help in the development of the tongue model, to Christophe Savariaux (ICP/GIPSA-lab) and Jacques Lebeau (Head of the plastic and maxillo-facial surgery department at Grenoble University Hospital) for theirs contributions to the analysis of patients after partial glossectomy and reconstruction.## V. References

1. Talmi YP. Quality of life issues in cancer of the oral cavity. Journal of Laryngology and Otology. 2002;**116**(10):785-90.
2. Deleyiannis FWB, Weymuller Jr. A, Coltrera MD. Quality of Life of Disease-Free Survivors of Advanced (Stage III or IV) Oropharyngeal Cancer. Head & Neck.1997; **19**:466-73.
3. Furia CLB, Kowalski LP, Latorre MRDO, Angelis EC, Martins NMS, Barros APB et al. Speech Intelligibility After Glossectomy and Speech Rehabilitation. Archives of Otolaryngology and Head Neck Surgery. 2001; **127**:877-883.
4. Perrier P, Savariaux C, Lebeau J, Magaña G. Speech production after tongue surgery and tongue reconstruction. Proceedings of the 14$^{th}$ International Congress of Phonetic Sciences, San Francisco, USA; 1999. Vol. 3, p. 1805-1808.
5. Savariaux C, Perrier P, Pape D, Lebeau J. Speech production after glossectomy and reconstructive lingual surgery: a longitudinal study. Proceedings of the 2$^{nd}$ International Workshop on Models and Analysis of Vocal Emissions for Biomedical Applications (MAVEBA). Firenze, Italy. 2001.
6. Bressmann T, Thind P, Uy C, Bollig C, Gilbert RW, Irish JC. Quantitative three-dimensional ultrasound analysis of tongue protrusion, grooving and symmetry: data from 12 normal speakers and a partial glossectomy. Clinical Linguistics & Phonetics. 2005; **19**(6-7):573-88.
7. Bressmann T. Speech adaptation to a self-inflicted cosmetic tongue split: perceptual and ultrasonographic analysis. Clinical Linguistics & Phonetics. 2006; **20**(2-3):205-10.
8. Lavallée S, Sautot P, Troccaz J, Cinquin P, Merloz P. Computer Assisted Spine Surgery : a technique for accurate transpedicular screw fixation using CT data and a 3D optical localizer. Journal of Image Guided Surgery. 1995; **1**(1):65-73.
9. Merloz P, Tonetti J, Eid A, Faure C, Lavallée S, Troccaz J, et al. Computer assisted spine surgery. Clin Orthop. 1997; **337**: 86-96.
10. Tonetti J, Carrat L, Lavallée S, Cinquin P, Merloz P, Pittet L. Ultrasound-based registration for percutaneous computer assisted pelvis surgery: Application to Iliosacral screwing of pelvis ring fractures. Computer Assisted Radiology and Surgery. 1997; 961 – 966.
11. Chabanas M, Luboz V, Payan Y. Patient specific Finite Element model of the face soft tissue for computer-assisted maxillofacial surgery. Medical Image Analysis. 2003; **7**(2):131-151.
15

12. Clatz O, Delingette H, Talos IF, Golby AJ, Kikinis R, Jolesz FA et al. Robust nonrigid registration to capture brain shift from intraoperative MRI. IEEE Trans Med Imaging. 2006; **24**(11): 1417-1427.
13. Ferrant M, Nabavi A, Macq B, Jolesz FA, Kikinis R, Warfield SK. Registration of 3-D intraoperative MR images of the brain using a finite-element biomechanical model. IEEE Transactions on Medical Imaging. 2001; **20**(12): 1384-1397.
14. Gérard J-M, Wilhelms-Tricarico R, Perrier P, Payan Y. A 3D dynamical biomechanical tongue model to study speech motor control. Recent Research Developments in Biomechanics. 2003;**1**: 49-64, Transworld Research Network.
15. Gérard J-M, Perrier P, Payan Y. 3D biomechanical tongue modelling to study speech production In: J. Harrington & M. Tabain Editors, Speech Production: Models, Phonetic Processes, and Techniques. Psychology Press: New-York, USA; 2006. p. 85-102
16. Buchaillard S, Pascal P, Payan Y. A 3D biomechanical vocal tract model to study speech production control: How to take into account the gravity? Proceedings of the 7th International Seminar on Speech Production, Ubatuba, Brazil; 2006. p. 403-10
17. Miyawaki K. A study on the musculature of the human tongue. Annual Bulletin of the Research Institute of Logopedics and Phoniatrics (Tokyo University). 1974; **8**: 22-50.
18. Netter FH. Atlas of Human anatomy. CIBA-GEIGY Corporation editor. 1999
19. Takemoto H. Morphological analyses of the human tongue musculature for three dimensional modelling. Journal of Speech, Language, Hearing Research. 2001; **44**: 95-107.
20. Wilhelms-Tricarico R. Development of a tongue and mouth floor model for normalization and biomechanical modelling. Proceedings of the 5th speech production seminar and CREST Workshop on models of speech production. 2000; 141-148. Kloster Seeon, Bavaria.
21. Badin P, Bailly G, Revéret L, Baciu M, Segebarth C, Savariaux C. Three-dimensional linear articulatory modeling of tongue, lips and face based on MRI and video images. Journal of Phonetics. 2002; **30**(3): 533-553.
22. Couteau B, Payan Y, Lavallée S. The Mesh-Matching algorithm: an automatic 3D mesh generator for finite element structures. Journal of Biomechanics. 2000; **33**(8); 1005-1009.
23. Feldman AG. Once more on the Equilibrium-Point hypothesis (λ model) for motor control. Journal of Motor Behavior. 1986; **18**(1):17-54.
24. Gérard J-M, Ohayon J, Luboz V, Perrier P, Payan Y. Non linear elastic properties of the lingual and facial tissues assessed by indentation technique. Application to the biomechanics of speech production. Medical Engineering & Physics. 2005; **27**: 884–892.
25. Duck FA. Physical Properties of Tissues: A Comprehensive Reference Book. Academic Press, London: 1990
26. Nesme M, Faure F, Payan Y. Hierarchical Multi-Resolution Finite Element Model for Soft Body Simulation. Lecture Notes in Computer Science. 2006; **4072**: 40-47.
27. Vogt F, Lloyd J, Buchaillard S, Perrier P, Chabanas M. Payan Y, et al.. Investigation of Efficient 3D Finite Element Modeling of a Muscle-Activated Tongue. Lecture Notes in Computer Science. 2006; **4072**: pp. 19-28.

**Table 1:** Maximum nodes velocities in cm/s on the surface of the tongue when the styloglossus is activated. Bracketed values give the velocity variation compared to the reference values.

|  | Apical | Palatal | Velar | Velopharyngeal | Pharyngeal |
|---|---|---|---|---|---|
| Reference | 18.74 | 13.77 | 19.46 | 18.29 | 6.98 |
| Flap (x 0.2) | 20.96 (+12%) | 14.91 (+8%) | 21.41 (+10%) | 20.03 (+10%) | 7.56 (+8%) |
| Flap (x 6.0) | 14.44 (-23%) | 10.68 (-22%) | 15.43 (-21%) | 14.86 (-19%) | 5.92 (-15%) |

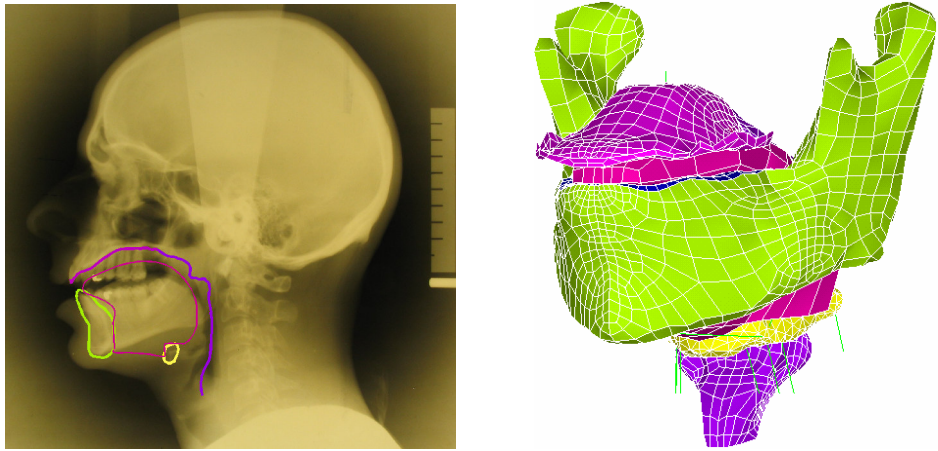

**Figure 1:** Final oral cavity geometry at rest. Left panel: Superimposition of the model midsagittal contours on X-ray at rest in the midsagittal plane. Right panel: the 3D tongue mesh in the whole oral cavity.

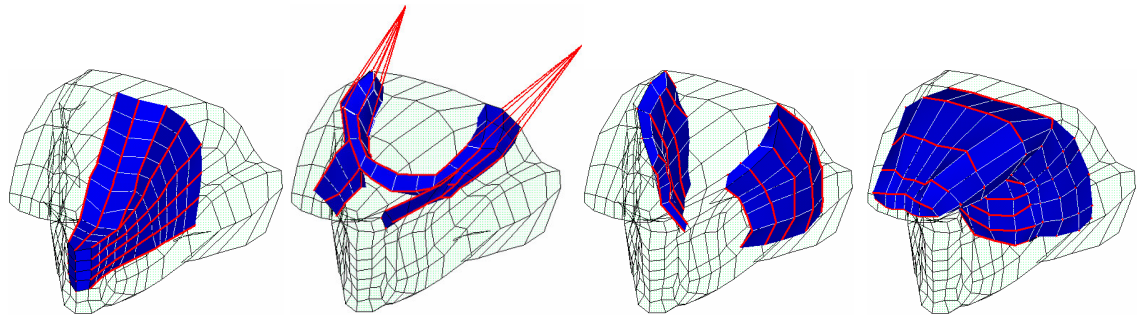

(a) Posterior genioglossus    (b) Styloglossus    (c) Hyoglossus    (d) Transversus

**Figure 2:** Representation of four muscles of the tongue (oblique view). The red lines represent the muscles macro-fibres.

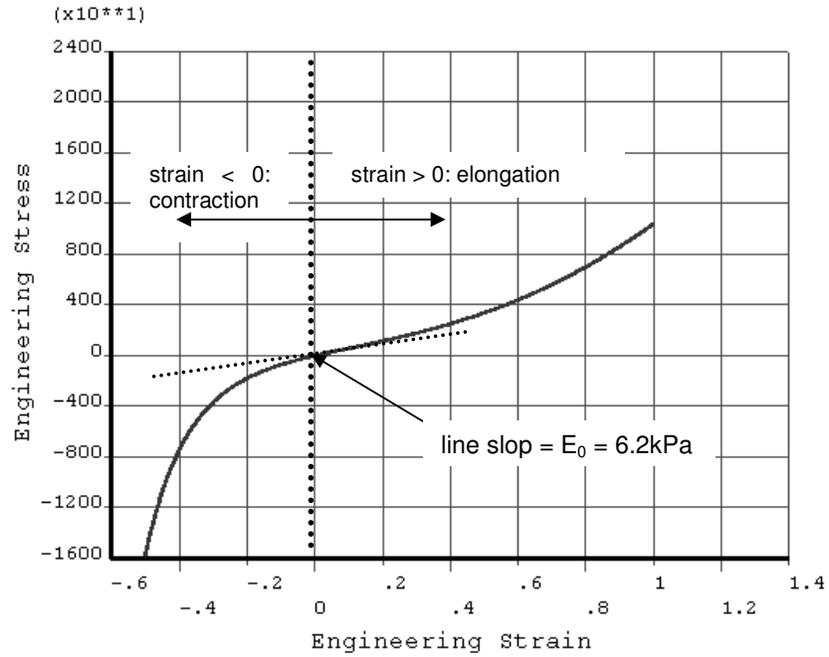

**Figure 3:** Stress/Strain constitutive law for passive tissues (Yeoh second order material with $C_{10}$=1037Pa and $C_{20}$=486Pa). The curve's tangent for a zero engineering strain gives the Young modulus at low strains.

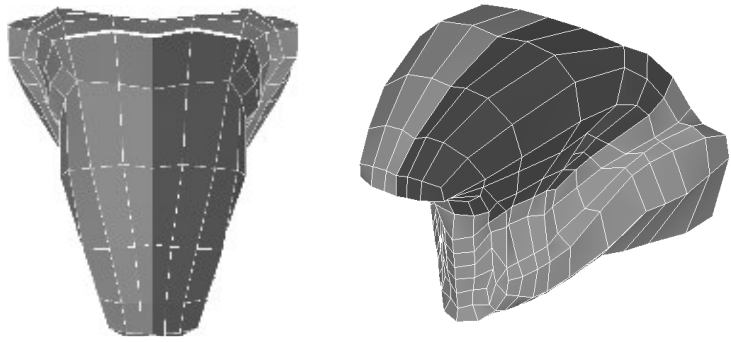

**Figure 4:** Simulation of a left hemiglossectomy of the tongue model. Left panel: transversal view of the mesh. Right panel: oblique view. Dark elements represent the resected tissues, and the white line symbolizes the line of the circumvallate papillae.

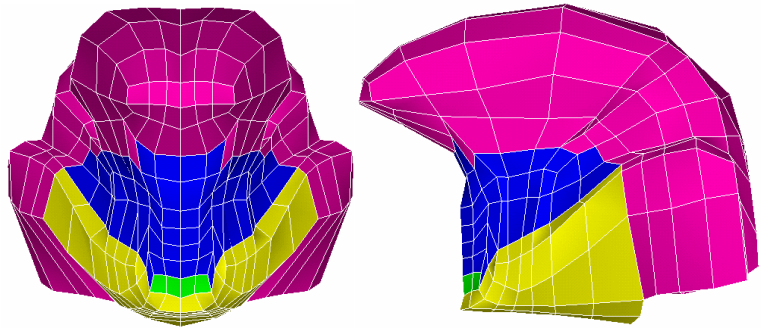

**Figure 5:** Simulation of mouth floor resection. Left panel: frontal view. Right panel: sagittal view. The resected part is represented in blue, green (geniohyoid muscle) and yellow (mylohyoid muscle).

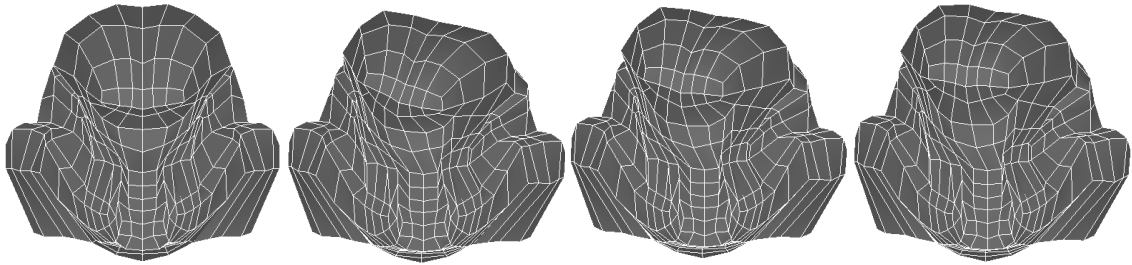

(a) Normal Model  (b) Small stiffness flap (x0.2)  (c) Medium stiffness flap(x1)  (d) High stiffness flap (x6)

**Figure 6:** Final tongue shape after activation of the styloglossus (frontal view). From the left to the right: the normal model and left-reconstructed models with a flap of increasing stiffness. The smaller the stiffness of the flap, the larger the asymmetry of the tongue shaping, characterised by the torsion of the apex and of the tongue dorsum, and with a deviation of the apex positioning from the midsagittal plane.

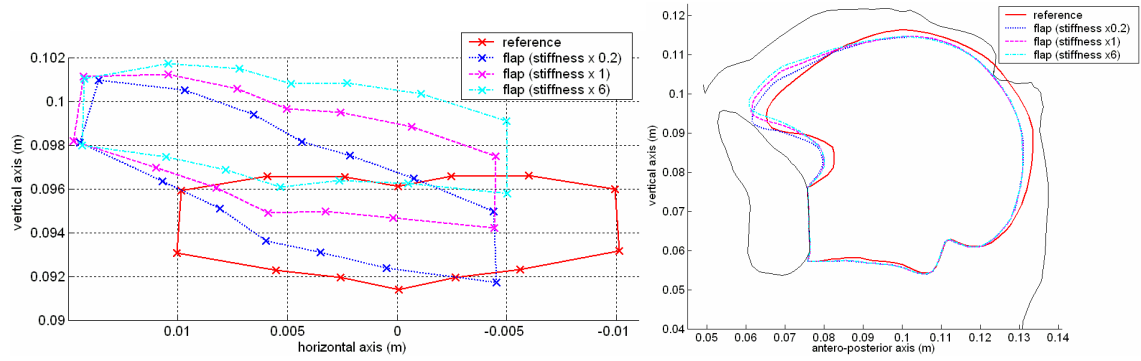

**Figure 7:** Final tongue shape after activation of the styloglossus. The left panel shows the position of the apex nodes in the frontal plane, and the right panel the shape of the tongue in the midsagittal plane. The solid line represents the normal model, the dotted line a left-reconstructed model with a small stiffness flap (x0.2), the dashed line a model with a medium stiffness flap (x1) and the dash-dot line a model with a high stiffness flap (x6).

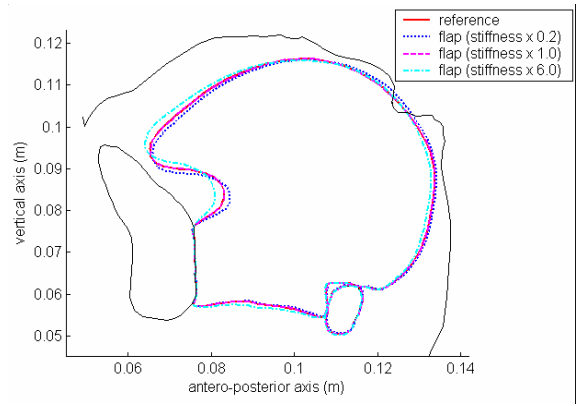

**Figure 8:** The solid line represents the normal model, the dotted line a reconstructed model with a small stiffness flap (x0.2), the dashed line a model with a medium stiffness flap (x1) and the dash-dot line a model with a high stiffness flap (x6).

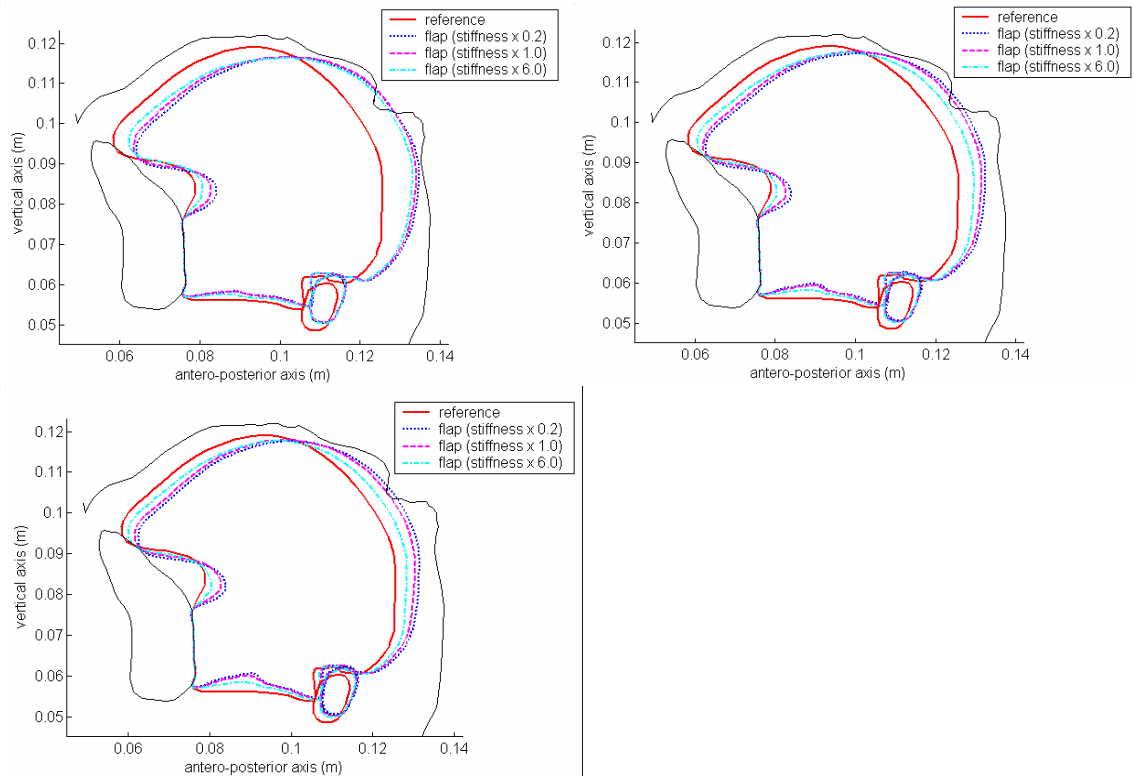

**Figure 9:** Co-contraction of the posterior genioglossus, styloglossus and transversus. Top-left hand corner: the genioglossus is not activated. Top-right hand corner: genioglossus is slightly activated. Bottom: genioglossus is highly activated. The solid line represents the normal model, the dotted line a reconstructed model with a small stiffness flap (x0.2), the dashed line a model with a medium stiffness flap (x1) and the dash-dot line a model with a high stiffness flap (x6).